\documentstyle[11pt]{article}
\begin{document}

\large \bf
Oscillatory
  Thickness Dependence of Magnetic Moments
 and\\ Interface-Induced Changes of the Exchange
       Coupling 
       \\ in Co/Cu and Co-Ni/Cu Multilayers
\vspace{1cm}

{\large S.~Krompiewski$^\dagger$, F.~S\"uss* and
 U. Krey*{\footnote{corresponding author}}}
\vspace{1cm}

\normalsize {\it $^\dagger$ Institute of Molecular Physics,
 P.A.N., Smoluchowskiego 17-19, 60-179 Pozna\'n, Poland \\
${}^*$ Institut f\"ur Physik II, Universit\"at Regensburg, 93040
 Regensburg, Germany}
 \vspace{1cm}

 Received ...........

\vspace{1cm}
\hrule \vspace {0.5cm} {\bf Abstract.}

We perform first-principles calculations  for the three multilayer systems
(100)-Co$_1$/Cu$_n$,
NiCo$_2$Ni/Cu$_n$ and Co$_4$Cu$_n$, and find from a
comparison of the results for system 2 and 3 that amplitude and phase
of  the exchange coupling
are sensitive to the magnetic-slab/nonmagnetic-spacer interface.
Moreover, we observe that
for the system 1 
 the averaged magnetic moment of the magnetic
slab oscillates with the spacer
thickness  similarly as the exchange coupling.
\vspace{0.5cm}\hrule \vspace {0.5cm}

It is now well known that in magnetic multilayers both the exchange coupling
($J$) as well as the giant magnetoresistance (GMR) depend not only
on the spacer
thickness but also on {\it kind} \cite{eins}, \cite{zwei} and {\it
thickness}
\cite{drei} -\cite{siebenA} of the magnetic slabs. According to
these results, the main
 effect of the  magnetic-slab thickness on the exchange coupling
oscillations is the appearence of an oscillatory phase shift without any
 substantial changes in amplitudes and periods. The { amplitudes} of the
oscillations
with the spacer thickness, in turn, depend on the
  {\it composition} of the ferromagnetic
layer, as was shown experimentally in \cite{acht,achtA}, where
Fe-Co-Ni/Cu fcc-(001) multilayers were studied.

The role of different interface atoms has
been studied in  detail first  by Parkin \cite{neun} who has shown
experimentally that the
GMR depends
critically on the contact interface monolayer. More recent systematic
studies of \cite{acht,achtA,zehn} suggest that this applies also to
the behaviour
 of $J$ as
far as its amplitudes and phases are concerned.
According to \cite{acht,achtA}, the exchange
coupling extrema
 of Co/Cu fcc-(001) superlattices get systematicly shifted when
the contact interface Co monolayer is replaced by either  Ni$_{0.5}$Co$_{0.5}$
or Ni  monolayers. The shift happens to be quite considerable, amounting to
roughly 1.5 \AA \, per extra electron per atom, see below. Additionally, a
strong reduction of the amplitudes is observed.

These experiments have motivated
us to apply
our  theoretical approach of \cite{sieben,siebenA}
 to study the {\it interface effect on the phase} of the oscillatory
exchange coupling.  (We use 8000 k-points, and our energies are
numerically accurate to 0.01 mRy, and in \cite{siebenA} we have shown
that our results in \cite{sieben,siebenA} agree with previous {\it
ab-initio}-calculations across Cu spacers,
e.g.~\cite{sieben1}-\cite{sieben4}).

Now  the difference in the number of electrons
per atom
of Ni and Co equals just $\Delta n=1$, and it has been stressed in
\cite{zwei} by employing a Friedel-Anderson-Caroli argument, that
 the relative phase shift of the results for NiCo$_{m-2}$Ni/Cu$_n$
with respect to
 Co$_m$/Cu$_n$ multilayers should scale with $\Delta n$, independent
of
the thicknesses involved. The observed phase shift of
  $\approx 1.5\,\AA$, see \cite{acht,achtA,zehn},
 is almost as big as one
 Cu-Cu interlayer spacing (1.8 $\AA$). This means e.g.~that the
presence of a minimum at a certain $n$ in the first-mentioned case
implies
the presence of a corresponding minimum of the second system at
$\approx n-1$
and {\it vice versa}.
  As already mentioned, also the amplitude of the
 oscillations is reduced considerably, namely
 by a factor 0.1 (0.3) for the first (second) antiferromagnetic
 peak, by replacing the interface Co atoms by Ni .

 We have been
 able to detect both effects, i.e.~the phase shift and the amplitude
reduction,
 by the accurate
 supercell spin-polarized linearized muffin-tin
orbitals method with the atomic sphere approximation (SP-LMTO-ASA),
which we use below,
\cite{zehnA}.
 A further point of our interest is the behaviour of the averaged
moment per atom of the magnetic slabs as a function of the composition
of the slab, particularly concerning the interface region, and of the
spacer thickness.  Since Cu itself is only slightly polarized
(typically $\sim 0.01 \mu_B$), and only at the interface, as we have
shown in \cite{siebenA}, the above-mentioned average moment is
essentially related to the total magnetic moment of a finite
 multilayer, if only the spacer thickness is varied.
 In fact, recent experiments,  \cite{elf}, have found
 for Ni/Au superlattices that this quantity, i.e.~the total magnetic
  moment divided by the  volume of
 the magnetic slabs, (and also
the Curie temperature of the  system)
 reveal oscillations with the spacer thickness.  In the
 present letter we find an analogous behaviour in Cu$_1$Co$_n$
 multilayers. 

To test the influence of the interface on the exchange coupling extrema
 we have
carried out  SP-LMTO-ASA supercell band
 calculations for Co$_4$/Cu$_n$ and NiCo$_2$Ni/Cu$_n$
fcc-(001) multilayers. Our method, when applied to the Co$_2$/Cu$_n$ and
Co$_n$/Cu$_2$
systems for $n=1,...,4$, \cite{sieben,siebenA},
 and for other systems \cite{zwoelf}, has already proved to
reproduce well both
 the overall behaviour of the exchange coupling with thickness-changes
of the ferromagnetic slab and/or of the spacer,
 and also the moment profiles across the multilayers
 (including a small, but significant spacer spin
polarization \cite{siebenA}, as already mentioned).
Although the exchange coupling amplitudes,
 when assumed to be proportional to the energy difference between the
parallel and antiparallel configurations, would be one order of
magnitude too large with our calculations,
 which unfortunately
seems to be
 typical, at present, for {\it ab-initio} calculations of the present kind
 in our field
  (see \cite{sieben} and comments
therein), it is encouraging that
 the crossover thicknesses between the two configurations as
well as the $J$ oscillation period lengths from our results do compare
 favourably with experiments.

In the following we show that the interface-induced phase shift is also well
reproduced by our method: Here one should note that to our knowledge
there have been no other ab initio studies
 of this interface effect on the exchange coupling phase so far;
attempts to explain it have been made only
 in terms of the Friedel-Anderson-Caroli
theory, in general  \cite{dreizehn}, and for some real systems of
our interest  in \cite{zwei,acht,achtA}.
In the latter  papers it has been shown that this theory works for a fcc-(001)
structure, whereas for the interpretation of the
fcc-(011) data an extension of the theory
is necessary.

We have studied structural models of Co$_m$/Cu$_n$ and NiCo$_2$Ni/Cu$_n$ with
$m=4$ and 1, and $n=1,...,6$. The atoms have been represented by spheres with
radii determined from the fcc-lattice constants $a$ of Ni, Co and Cu:
$a_{Ni}=3.524 $ \AA, $a_{Co}=3.548 $\AA\,  and
$a_{Cu}=3.615 $\AA\, ($a_{Co}$ is
calculated  from the hcp Co structure). The
superlattices have been constructed by placing successive atomic layers on
top of the basal one made of spheres representing copper
 with in-plane distances
equal to those of the bulk fcc-Cu. From these considerations the
following formula for the perpendicular interlayer spacings is found:
$    R_{i,j} = 0.5\cdot \{ \frac{1}{2}\cdot ( a_i + a_j )^2 - \alpha\cdot
 a_{Cu}^2 \}^{\frac{1}{2}}$,
where $\alpha = 1, 1.5$, and $\frac{2}{3}$ for  structures with
orientations (001), (011), and (111). Here we restrict
 ourselves to the (001) orientation, where
we get the following Cu-Cu, Co-Cu, Ni-Cu, Ni-Co,
and Co-Co interlayer spacings: 1.807, 1.774, 1.762, 1.727 and 1.740 \AA.

The main results of the present letter are presented in Fig.1 (a,b), where
the above mentioned average moment per atom of the magnetic slab
 in the stable configuration,
(a), and the total energy difference $\Delta E$
between antiparallel and parallel
configurations (per "ferromagnetic" supercell), (b), are plotted
vs.~the Cu spacer layers number.
The stable configuration is ferromagnetic (antiferromagnetic), when in Fig.~1b
$\Delta E$ is $>0$ ($<0)$. In Fig.~1a, for the Ni-Co/Cu
 system the ordinate-axis on the r.h.s.~of the plot should be used.
 It is easy to see from Fig.~1b that the presence of Ni at the
interface (full line) causes a shift of the exchange main minimum at
$n\approx 4.2$ of the NiCo$_2$Ni\,Cu$_n$ system by $\approx$ 0.8
Cu-monolayers to the r.h.s, i.e.~to $n=5$ for the Co$_4$Cu$_n$ system.
This is just what the experimentalists have measured, namely a shift
of $\approx 1.5 \,\, \AA$, see \cite{acht,achtA}.  Moreover, although our
{\it absolute} values of the exchange coupling amplitudes 
are again too high, the {\it ratio} of the
coupling strengths at the antiferromagnetic minima at $n\approx 4.2$ for
NiCo$_2$Ni/Cu$_4$ vs.~$n\approx 5$ for Co$_4$/Cu$_4$ is not
unreasonable: In our calculation, where there is only one Ni interface
monolayer, this ratio is $\approx 3$, whereas in the experiment, where the
Ni thickness is roughly three times as large, the ratio is $\approx
10$, see \cite{acht,achtA}.

To make this more quantitative, we have fitted the solid and dashed
lines in Fig.1b by the {\it ansatz} \begin{equation} \Delta
E(n)=\sum_{i=1,2}\, \frac{A_i}{n^{x_i}}\sin(2\pi
n/\lambda_i+\phi_i)\,, \end{equation} where for case (i), i.e.
 NiCo$_2$NiCu$_n$ (solid line), we set $x_1=2$, but $x_2=1$, while the
fit parameters are $A_1=0.786$ mRy, $\lambda_1 = 2.319$ monolayers
(ML), $\phi_1=-2.413$, $A_2= 0.47$ mRy, $\lambda_2 = 7.899$ ML and
$\phi_2= 0.804$. For the dashed line, i.e.~the Co$_4$Cu$_n$ system, a
similar fit has been obtained, with $x_i=2$: In this case,
(ii), the fit parameters are $A_1=3.13$ mRy, $\lambda_1=2.63$ ML,
$\phi_1=0.693$, $A_2=12.1$ mRy, $\lambda_2=32.6$ ML,
$\phi_2=2.60$. Thus the short wavelength $\lambda_1$ is hardly
influenced by the Ni substitution, and the corresponding phase shift
is $\phi_i^{(i)}-\phi_1^{(ii)}=-3.11$. Thus, from the sinus function
with the
 short wavelengths $\lambda_1$ one easily derives that the minimum of
$\Delta E$ for NiCo$_2$NiCu$_n$ (solid line) at $n=5$ corresponds to a
minimum of $\Delta E$ for Co$_4$Cu$_n$ at $n\approx 4.3$, in agreement
with the experiment.

\hglue 7 pt
Another noteworthy point is the behaviour of the average magnetic moment
per atom of the magnetic slab: In the
cases of Co$_4$Cu$_n$ and NiCo$_2$NiCu$_n$
 it changes only slightly  with $n$, and there (dashed line and solid line)
 the change is not
correlated with the behaviour of the exchange coupling. In
contrast, for  Co$_1$/Cu$_n$ the effect is much larger,
 and the average
 moment
 follows rather strictly the behaviour of the coupling (dotted line).
However,
it is also remarkable that for NiCo$_2$NiCu$_n$ the average moment is
considerably reduced from $1.057\,\mu_B $ for $n=1$ to $1.02 \,\mu_B$
for
$n \ge 2$.
In  Fig.~2 we find that the main part ($\sim 2/3$)
of the last-mentioned reduction is due to Ni, which is
rather sensitive
to changes of position and neighbourhood effects, see
\cite{zwoelf}, although the Co moments, too, are weakened roughly by 2/5
 of
the reduction of Ni moments. 
  In contrast,
 in a Co$_4$/Cu$_n$ multilayer the internal Co-Co
 coupling seems to make the Co system rather stiff with respect to changes
 of the spacer thickness $n$, see the dashed curve in Fig.~1a.
For  Co$_1$Cu$_n$, this Co-Co ''stiffness'' is of course not present, which may
be the reason for the fact that in this case the
 dotted moment
 curve for Co$_1$Cu$_n$ in Fig.~1a
 follows rather closely that of the exchange coupling in Fig.1b.

 In conclusion, it has been shown by accurate  SP-LMTO-ASA
 band calculations that
{\it phase} and {\it amplitude}
of the oscillatory exchange coupling depends
 strongly on the contact interface
magnetic layer. We have found also that there may be superlattices,
 which reveal
 oscillations  of the magnetization of the magnetic
slabs with the spacer thickness, which are strongly similar to
 the oscillations of the
exchange coupling. This seems to be the
 case when the magnetic slabs are only one monolayer thick, or perhaps
more generally when the exchange
self-coupling within the magnetic slab is small.

{\bf Acknowledgements}

This work has been carried out under the grant no. 2 P 302 005 07 (SK), and
the bilateral project DFG/PAN X.083 (UK and SK). We also thank the J\"ulich,
Munich,
Regensburg and Pozna\'n Computer Centres for computing time.


\newpage

{\Large \bf Figure Captions}
\vspace{1cm}

Fig.1: The average magnetic moment $\langle \mu\rangle$
 per atom of the magnetic slab, (a),
 and the total energy
       difference $\Delta E$ per ferromagnetic unit cell
       between the parallel and
       antiparallel configurations, (b), are plotted vs.~the Cu-spacer
        thickness, for
       the following superlattices: NiCo$_2$Ni/Cu$_n$,
       (solid line), Co$_4$/Cu$_n$ (dashed line) and Co$_1$/Cu$_n$ (
       dotted line). The arrows indicate the phase shift upon
       replacing  Co at the interface layers in Co$_4$/Cu$_n$ by Ni.
       Note that for the
NiCo$_2$Ni/Cu$_n$ multilayers the right ordinate axis applies.
\vglue 0.5 truecm
Fig.2: The Co and Ni moments in NiCo$_2$Ni/Cu$_n$ multilayers are plotted
 against $n$. Note the separate ordinate axes: Left ordinate
 for Co moments, right
one for Ni.
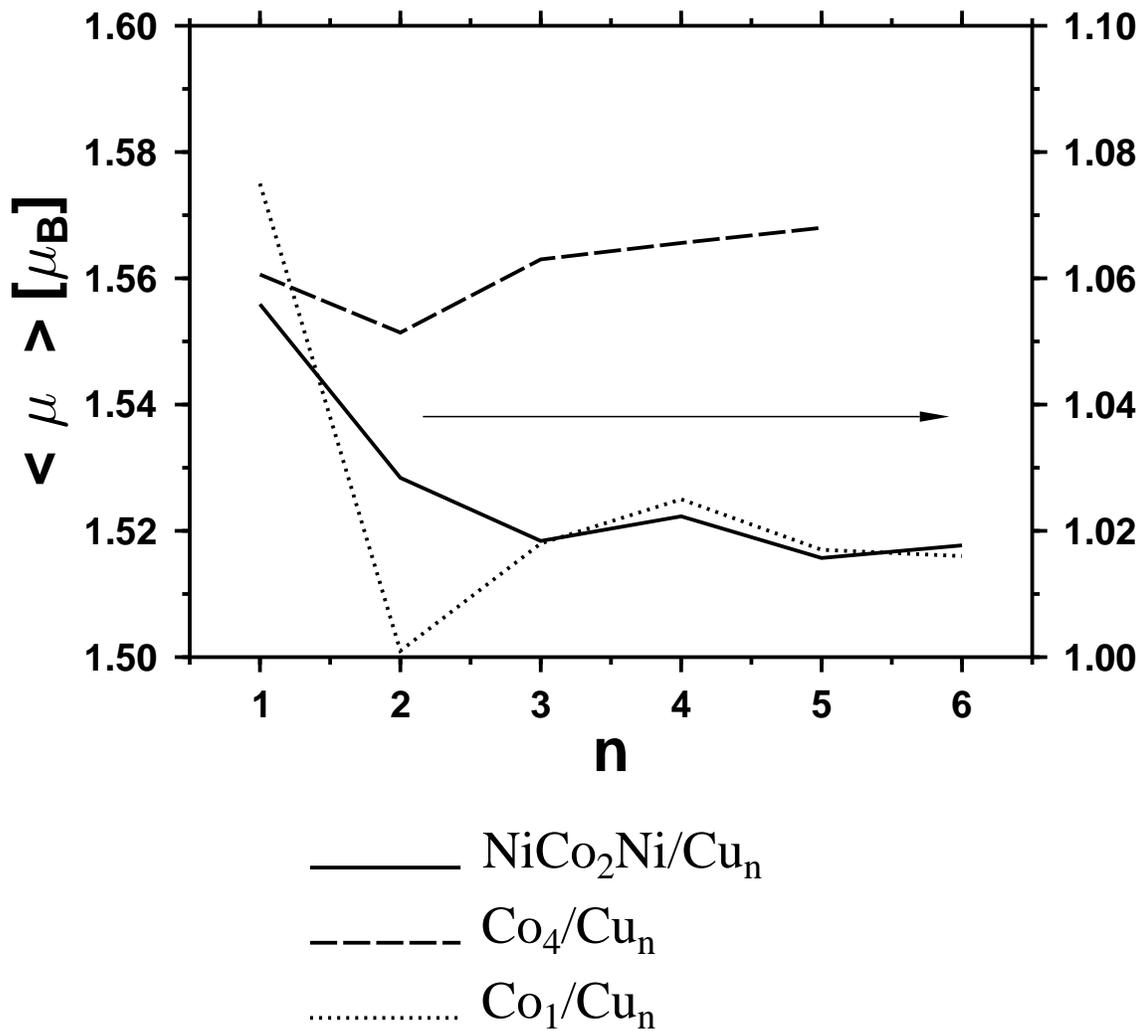
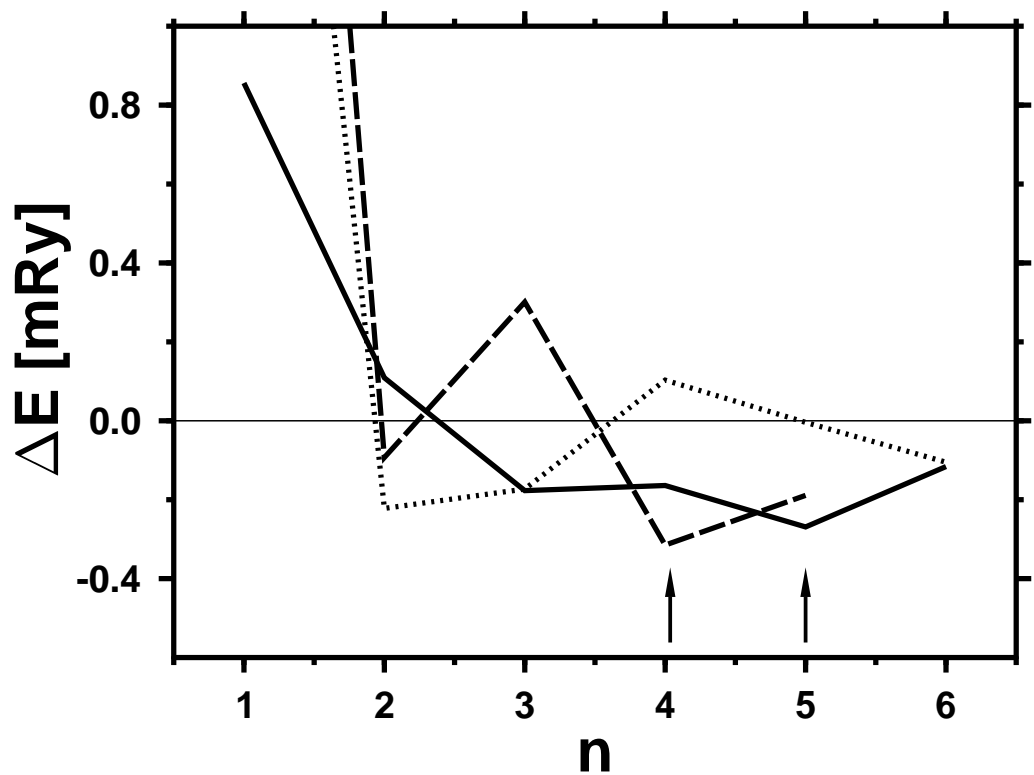
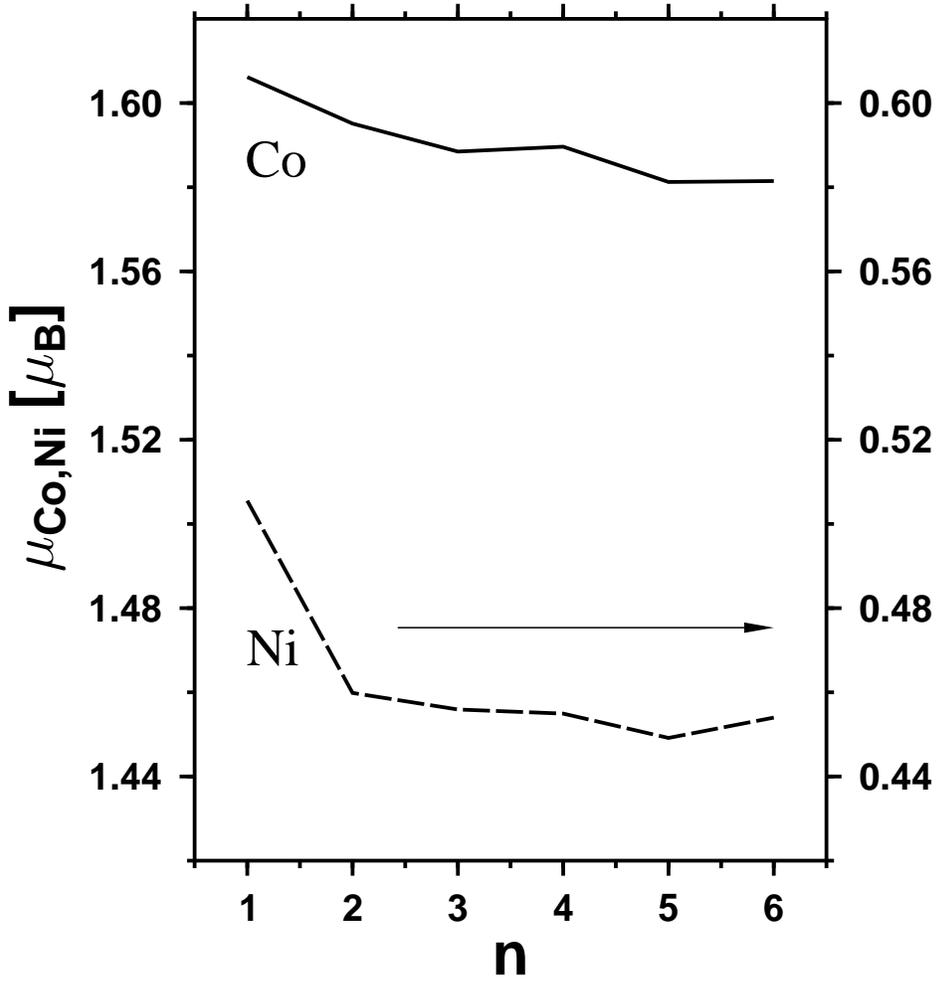
\end{document}